\documentclass[sn-standardnature, iicol]{sn-jnl}
\usepackage{xcolor}
\usepackage{color, soul}

\usepackage{marginnote}
\let\oldequation\equation
\let\oldendequation\endequation

\renewenvironment{equation}
  {\linenomathNonumbers\oldequation}
  {\oldendequation\endlinenomath}

\jyear{2024}%

\theoremstyle{thmstyleone}%
\theoremstyle{thmstyletwo}%

\theoremstyle{thmstylethree}%

\begin{document}

\title[Article Title]{
Silicon Photonic Microresonator-Based High-Resolution Line-by-Line Pulse Shaping
}


\author*[1]{\fnm{Lucas} \sur{M. Cohen}} \email{\small cohen26@purdue.edu}

\author*[1]{\fnm{Kaiyi} \sur{Wu}} \email{\small wu1871@purdue.edu} 

\author[1]{\fnm{Karthik} \sur{V. Myilswamy}}

\author[1]{\fnm{Saleha} \sur{Fatema}}

\author[2]{\fnm{Navin} \sur{B. Lingaraju}}

\author[1]{\fnm{Andrew} \sur{M. Weiner}}

\affil[1]{\small\orgdiv{School of Electrical and Computer Engineering}, \orgname{Purdue University}, \orgaddress{\city{West Lafayette}, \postcode{47907}, \state{IN}, \country{USA}}}

\affil[2]{\small\orgdiv{The Johns Hopkins University Applied Physics Laboratory}, \orgaddress{\city{Laurel}, \postcode{20723}, \state{MD}, \country{USA}}}

\abstract{

Optical pulse shaping stands as a formidable technique in ultrafast optics, radio-frequency photonics, and quantum communications. While existing systems rely on bulk optics or integrated platforms with planar waveguide sections for spatial dispersion, they face limitations in achieving finer (few- or sub-GHz) spectrum control. These methods either demand considerable space or suffer from pronounced phase errors and optical losses when assembled to achieve fine resolution. Addressing these challenges, we present a foundry-fabricated six-channel silicon photonic shaper using microresonator filter banks with inline phase control and high spectral resolution. 
Leveraging existing comb-based spectroscopic techniques, we devise a novel system to mitigate thermal crosstalk and enable the versatile use of our on-chip shaper. Our results demonstrate the shaper's ability to phase-compensate six comb lines at tunable channel spacings of 3, 4, and 5 GHz. Specifically, at a 3 GHz channel spacing, we showcase the generation of arbitrary waveforms in the time domain. This scalable design and control scheme holds promise in meeting future demands for high-precision spectral shaping capabilities.

}

\keywords{integrated photonics, line-by-line pulse shaping, optical arbitrary waveform generation}



\maketitle

\begin{figure*}[!t]%
\centering
\includegraphics[width=0.9\textwidth]{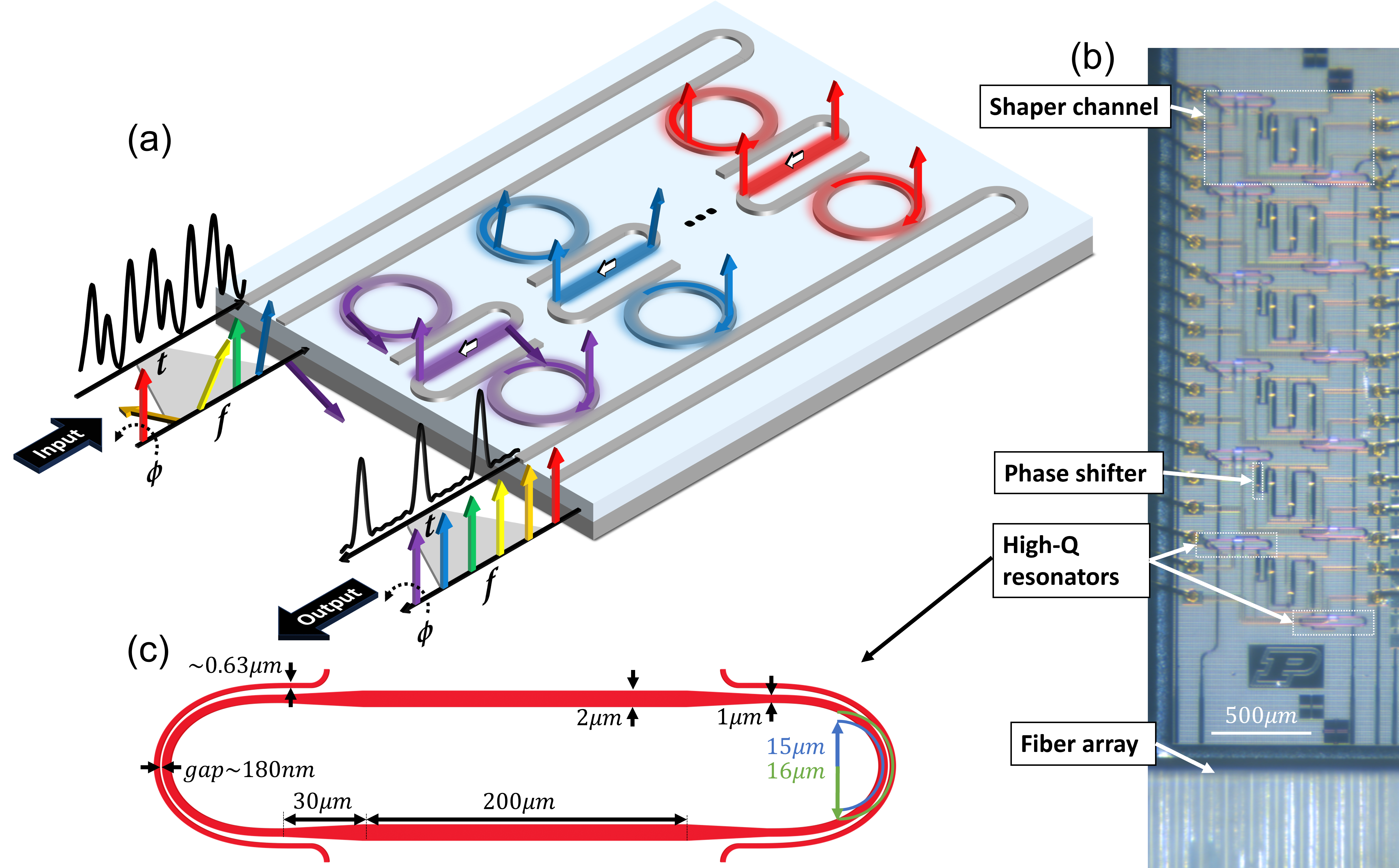}
\caption{(a) Graphic illustration of a six-channel microresonator-based spectral shaper, and a conceptual illustration of line-by-line pulse shaping. An input optical frequency comb with random relative phases between comb lines generates a distorted waveform in the time domain. The spectral shaper chip can be programmed to align the phases and produce transform-limited pulses at the output. (b) Top-down microscope image of the fabricated system. (c) Zoomed-in schematic of a single resonator device used in our shaper. }\label{fig1}
\end{figure*}


By controlling the amplitude and phase of spectral lines over a coherent, broadband optical spectrum, 
optical pulse shaping has been widely used in applications for ultrafast optical arbitrary waveform generation~\cite{Cundiff2010,weiner2011ultrafast,Jiang2007,Willits:12,liao2015arbitrary,divitt2019ultrafast}, wideband radio frequency (RF) photonics~\cite{lin2005photonic,huang2008spectral, khan2010ultrabroad,zhang2015silicon,Wu2018a,DeChatellus2018,schnebelin2019programmable,Pan2020a,redding2022high}, and manipulation of frequency-encoded photonic quantum information~\cite{lu2018quantum,kues2019quantum,lu2023frequency}. Fourier transform pulse shapers based on bulk optics comprise a pair of spectral dispersers with spatial light modulators between them. The first spectral disperser scatters, or demultiplexes, a broadband field such that individual spectral components are routed to unique light-modulating elements. Following modification of the amplitude and phase by the spatial light modulator, the spectral components are recombined, or multiplexed, at the second spectral disperser~\cite{weiner2011ultrafast}. 
Commercially available pulse shapers rely on bulk spectral dispersers
and liquid-crystal-on-silicon (LCoS) spatial light modulators to control the amplitude and phase of the input signal~\cite{weiner2000femtosecond,coherentWaveshaper}. In laboratory settings, it is common to replace or pair diffraction gratings with virtually imaged phase arrays 
to enhance the spectral resolution~\cite{Xiao2005,lee2005programmable,Lee2023}.

While off-the-shelf equipment is well-suited for femtosecond, high-repetition-rate optical waveform manipulation~\cite{Jiang2007}, the technology has severe limitations when one requires finer control over the optical spectrum as is the case with low-repetition rate optical pulse shaping or high resolution spectrometry~\cite{Willits:12}. In particular, few- or sub-GHz addressability is invariably associated with a significant loss penalty and also requires an increase in the system size, which is prohibitive outside of the laboratory environment.

High resolution optical pulse shaping has proven invaluable across diverse quantum applications.
The technology would enable straightforward manipulation and characterization of frequency-encoded photonic quantum information by reducing the separation between addressable modes in the computational space~\cite{myilswamy2023time,lu2023frequency}.
In addition, it facilitates fast waveform modulation to generate voltage standards traceable to fundamental quantum limits~\cite{Lee2023}, as well as precise control of optical transitions in molecular ions or atoms~\cite{Ma2020,Stollenwerk2020}. 
Furthermore, high spectral resolution is vital in microwave photonics for generating and manipulating longer waveforms~\cite{schnebelin2019programmable,redding2022high}. Fine-grained control over the spectral amplitude also has tremendous value in applications where control over the spectral phase is not required. This is particularly true in spectroscopy, where high resolution spectral filtering has been used with Brillouin light-scattering techniques for non-invasive biological imaging~\cite{Meng2016,Kabakova2022} and to characterize stellar illumination in astrophotonics applications~\cite{Gatkine2019,jovanovic20232023}.

As an enabling technology, photonic integrated circuits (PICs) offer a low size, weight, and power (SWaP) chip-scale platform for pulse shaping. 
The silicon-on-insulator (SOI) platform especially benefits from mature CMOS process compatibility to enable a rich component ecosystem with the capability for volume production of complex PICs.
Numerous on-chip shapers have been shown employing arrayed waveguide gratings (AWGs)  as spectral dispersers~\cite{fontaine200732,FONTAINE20113693,soares2011monolithic,metcalf2016integrated,Feng2017a}. 
However, to achieve a sub-GHz spectral resolution, AWG-based spectral dispersers require an excessively large ($\sim$ cm$^2$) device footprint which leads to increasing phase errors
on platforms like SOI~\cite{gehl2017active}. Also, AWGs do not offer tunability resulting in channel bandwidths and spacings that are fixed. To address these issues, shapers with a microresonator-based spectral disperser, whose channel bandwidth and minimum spacing depend on the quality (Q) factor of the microresonators, have been explored~\cite{agarwal2006fully, khan2010ultrabroad, wang2015reconfigurable}. 
Although previous demonstrations have used microresonators with a low Q factor (linewidth $\sim 8$~GHz or greater), microresonators on SOI can be designed to be extremely wavelength selective with a high Q factor of $> 10^{6}$~\cite{Onural2021, Cohen2023} and can be realized with a small device footprint of $\sim 10^{-2}$ mm$^2$.
Their center frequency can be tuned via the thermo-optic (TO), free-carrier dispersion, or other~\cite{Rios2022} dispersive effects. TO tuning is common because it offers simple design and fabrication, a wide tuning range, and does not fundamentally induce optical loss. 
However, thermal crosstalk introduces challenges in the large-scale control of such shapers. 

\begin{figure*}[!t]%
\centering
\includegraphics[width=1\textwidth]{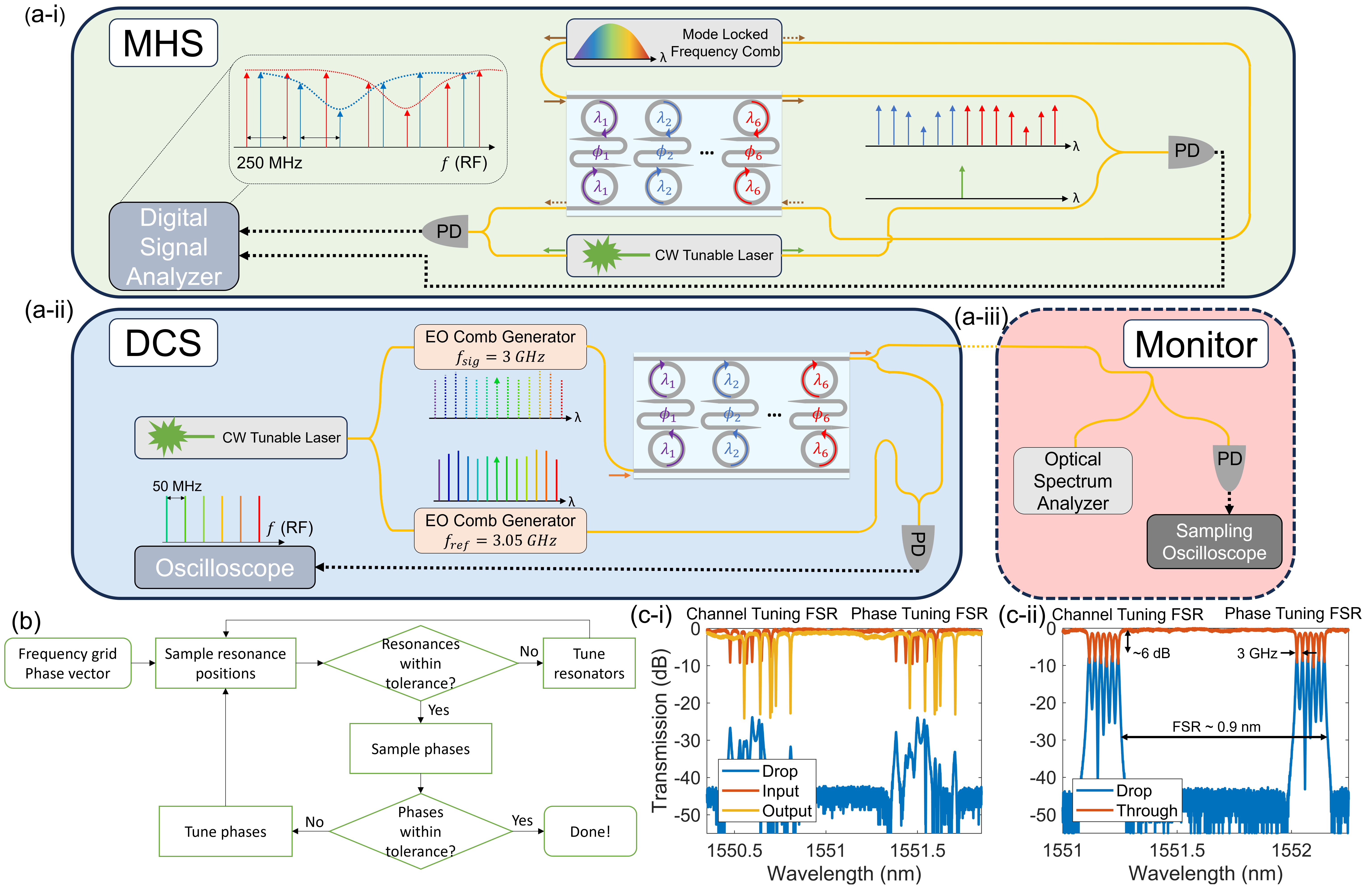}
\caption{
High-level schematic diagram of experimental setups for (a-i) Multi-Heterodyne Spectroscopy (MHS), (a-ii) Dual-Comb Spectroscopy (DHS), for measuring the channel frequencies and phases, respectively, and (a-iii) the Waveform Monitor. Yellow and dashed black lines indicate optical and RF connections, respectively. Brown (solid and dashed) and green arrows in (a-i) follow directions of propagation for the MLFC and CW laser, respectively. Orange arrows in (a-ii) follow the signal comb. Arrows on the resonators within the SiP shaper follow the direction of propagation for each method. For clarity, the SiP shaper is drawn here with optical I/O on both sides of the chip; in the experiment, optical I/O is on a single side, as depicted in Fig.~\ref{fig1}(a). MHS and DCS measurements are run at the same time by adding circulators and DWDM filters. PD: Photodiode. (b) Flowchart of our shaper control scheme. (c-i) Transmission spectrum of the as-fabricated spectral shaper and (c-ii) with the shaper programmed to compress six lines from a $3$~GHz EO comb. In both cases, the measurement is from a swept laser source and the two FSRs for MHS and DCS measurement are shown. Transmission is normalized to coupling loss to/from the chip, which we measure to be $\sim 3.5$~dB/facet.
}\label{fig2}
\end{figure*}


In this paper, we design, fabricate, and optimize
high resolution ($\sim 900$~MHz linewidth) microresonator filter banks with inline phase control for narrowband pulse shaping. 
We employ a novel method of accurately aligning such shapers onto arbitrary frequency grids with any prescribed phase. In particular, we combine multi-heterodyne spectroscopy (MHS)~\cite{urabe2013multiheterodyne} and dual-comb spectroscopy (DCS)~\cite{Coddington2016} to sample channel frequencies and phases over short acquisition times and with high spectral resolution.
This allows real-time monitoring of the system response and accurate tuning of the shaper elements in the presence of thermal crosstalk or environmental disturbances. We employ our method with a six-channel silicon photonic (SiP) device with sub-GHz channel linewidths programmed to operate on a $3$~GHz grid to synthesize arbitrary user-defined optical waveforms in the time domain. 
We emphasize the tunability of our shaper by extending the waveform demonstrations to $4$ and $5$~GHz frequency grids to phase-compensate residual phases on individual lines of an electro-optic (EO) frequency comb.  
These demonstrations mark a significant step toward realizing narrowband filtering and high resolution shaping functionality in a scalable manner.

\vspace{\baselineskip}
\noindent{\textbf{Results}}

\noindent{\textbf{SiP shaper design}} 

Figure~\ref{fig1}(a) is a high-level concept illustration of our SiP microresonator-based spectral shaper, and Fig.~\ref{fig1}(b) displays its microscope image. A common bus waveguide routes an input broadband optical field to a bank of resonators. The input resonator bank downloads slices of the input spectrum and routes them through phase-shifting elements before finally uploading the slices onto a common output bus via the second resonator bank. 
Transmission spectra of both the input and output resonator banks can be accessed. 
The shapers are fabricated through AIM Photonics~\cite{Fahrenkopf2019} and wirebonded to a custom interposer and printed circuit board (PCB). 

Resonators in our device [diagram in Fig~\ref{fig1}(c)] are designed using multimode waveguides. 
Operating these waveguides in the fundamental mode substantially diminishes the predominant loss mechanism of field-sidewall overlap in SiP waveguides, thereby ensuring low-loss performance~\cite{Zhang2020a,Cohen2022}.
The resonators contain 
200$\mu$m of straight 2$\mu$m-wide waveguides
which taper linearly over 30$\mu$m to a width of 1$\mu$m at the coupling sections. This permits higher rates of coupling to the resonator, reducing the drop-port loss of the filter. To suppress higher-order mode excitation, the resonator and the bus waveguides follow an Euler curve~\cite{Jiang2018}. The curve in the resonator reaches a minimum bending radius of 15$\mu$m, and the 
bus waveguides are designed to be phase-matched at the midpoint of the curve, where the local bending radius, width, and gap are  16$\mu$m, 0.64$\mu$m, and 0.18$\mu$m, respectively. A single resonator occupies a compact footprint of only 0.045$\times$0.33~mm$^{2}$.
Each resonator has a measured $3$~dB linewidth, drop-port loss, and free spectral range (FSR) of $\sim 1.1$~GHz, $\sim 2.5$~dB, and $\sim 115$~GHz, respectively. Thermo-optic phase shifters, comprised of doped silicon slabs functioning as resistive heating elements, are placed between resonators in a channel and also embedded within each resonator for tuning the center frequency. Variable optical attenuators are also placed in-line within a channel for amplitude tuning, although they are not used here. When resonators on the input and output sides are aligned to a common resonance mode, the composite shaper channel linewidth and drop-port loss are $\sim 900$~MHz and $\sim 6$~dB. Progress in fabrication processes compatible with CMOS technology may enable additional reduction in these quantities in the near future~\cite{fanto,9220770}. 


The input and output resonators within any channel must have their center frequencies well-aligned to allow for efficient transmission of a spectral slice. Variation in silicon device layer thickness and sidewall roughness across a die leads to a starting resonant frequency mismatch between filters in a channel. Microheaters allow for thermal tuning of the resonant frequencies, but the simultaneous heating of devices leads to compounding thermal crosstalk making precise alignment challenging. For these reasons, and to exploit the tunability of the resonators, a robust control method must be employed to align resonators within the shaper to specific frequency positions.

\noindent{\textbf{MHS/DCS programming method}}

Figure~\ref{fig2}(a-i) illustrates the experimental setup for the MHS measurement technique. We use a continuous wave (CW) laser and a stabilized mode-locked frequency comb (MLFC) for this purpose. We note that although the concept of MHS and multi-heterodyne beat generation also includes the beating between two frequency combs (i.e., DCS, which we adopt for phase measurement discussed later), we still designate this measurement as MHS to differentiate between this and the latter method.  
A portion of the spectrum from 
the MLFC is sent through bus waveguides on both the input and output sides of the shaper as indicated in Fig.~\ref{fig2}(a-i). The coupling between the bus waveguides and resonators leads to a significant ($\sim 10$~dB) frequency-dependent extinction as measured from the output of the bus waveguides, indicating the resonance frequencies of the resonators.
In this way, the frequency response of the resonator bank is encoded onto the amplitude of the comb lines.
By combining the spectrally modified comb exiting the bus waveguides with a CW laser operating at a fixed wavelength and beating on a high-speed photodetector (PD), the optical response of the resonator bank can be downconverted to the RF range after measurement by an oscilloscope and fast-Fourier transform (FFT). The optical spectrum is revealed as amplitude-modified tones spaced at the repetition rate of the comb ($f_{\rm rep}=250$~MHz). This MHS measurement enables multiple optical spectra to be measured at the same time (limited by the number of oscilloscope channels) and supports high rates of measurement, with a single measurement requiring an acquisition time on the order of $\sim \mu$s. 
Here, we employ this method to measure the optical spectrum of both resonator banks simultaneously (example spectra can be found in the Supplementary Information).
By varying the resonator drive currents incrementally and processing the RF spectrum each time, the resonance frequencies of both resonators within each channel can be tracked via peak finding methods and tuned to arbitrary frequency positions (relative to the CW laser and with a precision of the MLFC repetition rate). 
In our experiments, we align shaper channels to a uniformly spaced frequency grid.

To measure the spectral phase applied to each channel, we adopt a DCS method with two EO combs [schematic in Fig.~\ref{fig2}(a-ii)] generated using the same pump laser.
The signal comb is driven with a frequency equal to the spacing of the shaper channels ($f_{\rm sig}$) and the reference comb is driven at a slightly different frequency ($f_{\rm ref}=f_{\rm sig}+\Delta f$). The two oscillators driving the combs are synchronized with each other.
Signal comb modes on the high-wavelength side of the pump are sent through the shaper to accumulate the phase applied to each channel. The signal comb exiting the chip through the output port is combined with the reference comb and sent to a 
balanced PD. Heterodyne beats at harmonics of the dual-comb difference frequency ($\Delta f$) are generated. The phase difference between the signal and reference comb can be measured by acquiring the generated electronic signal using an oscilloscope and sampling the phase at beat note frequencies after an FFT. Similar to MHS, the DCS measurement can be executed at MHz rates~\cite{Duran2015}. More details can be found in Methods and Supplementary Information.

\begin{figure*}[!t]%
\centering
\includegraphics[width=0.9\textwidth]{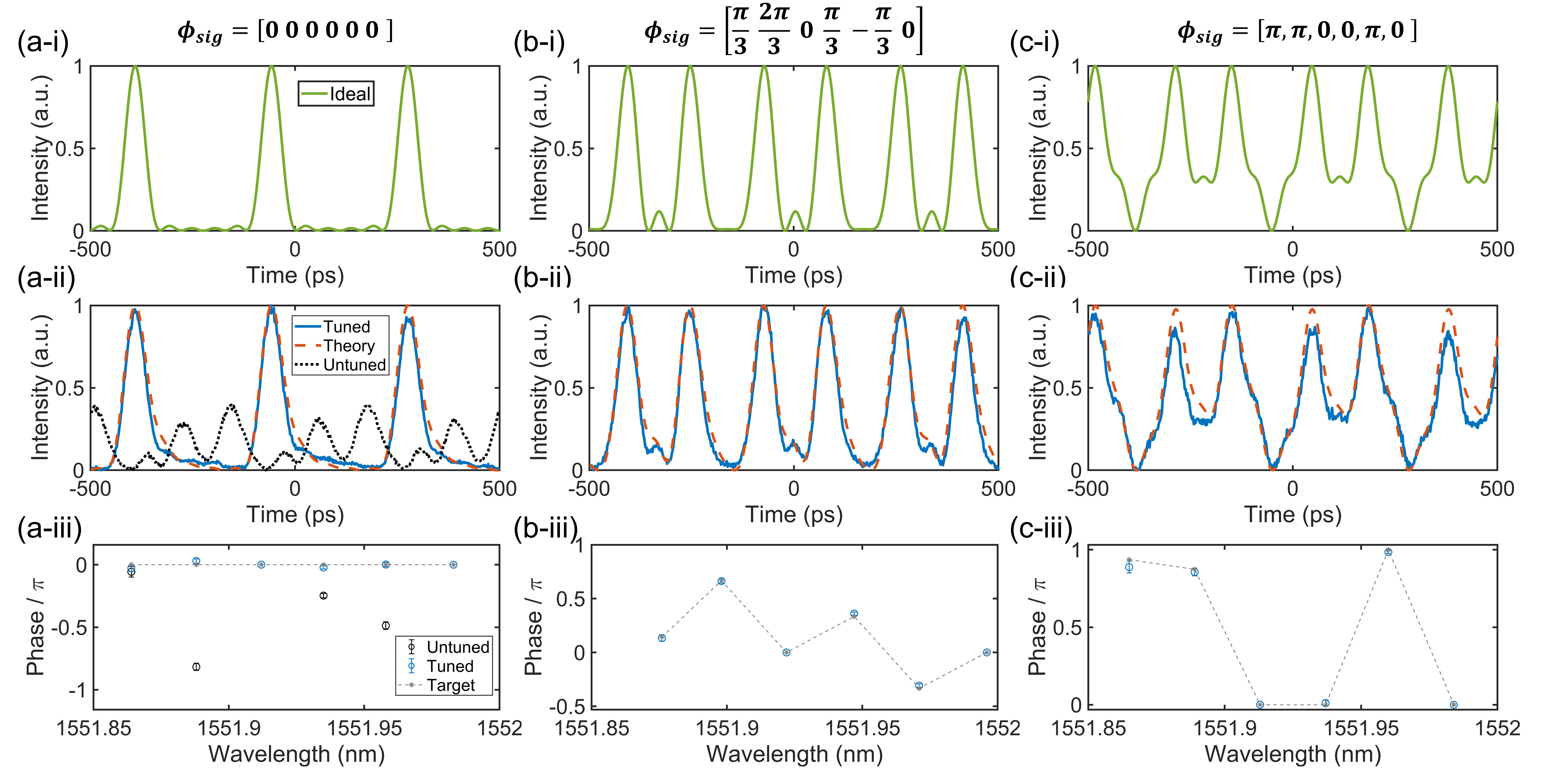}
\caption{\textbf{Optical Arbitrary Waveform Generation} (a,b,c-i) Ideal target waveforms in the time domain (green traces) and signal comb spectral phase (top) for the compressed, Talbot, and forked state, respectively. (a,b,c-ii) Measured waveforms (blue trace) and the ideal waveforms convolved with the measurement system impulse response (dashed orange traces) for the compressed, Talbot, and forked states, respectively. An additional waveform is shown in (a-ii) (black dotted trace) with an untuned phase and phase shifters driven to induce $\sim \pi$ phase. (a,b,c-iii) Target (gray dashed) and measured (blue scattered with error bars) signal comb spectral phase for the compressed, Talbot, and forked states, respectively. An additional measurement is shown in (a-iii) (black scattered with error bars) for the untuned phase state, with phase shifters driven to induce $\sim \pi$ phase. Figure legends in (a-i,ii,iii) are shared amongst subplots in a row, and the untuned state is only shown in (a-ii,iii). All theoretical waveforms take into account the optical power of each comb line as measured by OSA. Error bars (standard deviation) in the phase plots are difficult to see. These errors always $< 0.15$~rad (see Methods and Supplementary Information), and are on average $\sim 0.05$~rad. See Fig.~\ref{fig4}(a,b,c-iii) for a clear example.}\label{fig3}
\end{figure*}

The MHS and DCS measurements can be performed simultaneously in a bi-directional manner by employing circulators and dense wavelength division multiplexing (DWDM) filters. The periodic spectral response of the resonators permits a particular FSR to be used for MHS while an adjacent FSR is used for DCS phase measurement.
We develop a Python-based control routine that takes as input a frequency grid and relative phase vector and iteratively performs MHS and DCS to track and align shaper channels onto the frequency grid and drive phase shifters to the appropriate relative phase values. The phase shifters are pre-driven with a power slightly higher than that to induce a $\pi$-phase shift at the start so that the drive can be increased or decreased to reach the desired phase. Likewise, the resonators are pre-detuned in a non-overlapping manner to simplify the alignment. The sequence of operations within the routine is shown in the flowchart of Fig.~\ref{fig2}(b). First, the MHS method is used to repeatedly measure the center frequencies of both resonator banks and increment the respective resonator drive signals to align channels into the desired frequency positions. Once this measurement indicates channels are tuned correctly, the DCS signal is employed at the adjacent FSR, and the relative phase on each line after propagation through the shaper is computed. Phase shifters within each channel are then driven in small increments towards the target phase values. If this causes the resonators within a channel to be detuned from their target frequency positions due to thermal crosstalk, the MHS method with electronic feedback is repeated to realign them. Then the phases are again measured and incremented using DCS. This process repeats until shaper channels are aligned to the grid and phase shifters induce the target phase on each line. 
In this way, channels can be well aligned onto a frequency grid (with error $\sim f_{\rm rep}$ of the MLFC), the phases on each line can be accurately measured and set (with standard deviation $<0.15$~rad), and thermal crosstalk effects can be compensated for. More details can be found in Methods and Supplementary Information. 
The through and drop port transmission spectrum of the device before and after using this method to align channels to a 3 GHz grid with uniform phases is shown in Fig.~\ref{fig2}(c-i) and Fig.~\ref{fig2}(c-ii), respectively. Note the FSRs used for both MHS and DCS are shown.

\noindent{\textbf{Optical arbitrary waveform generation}}

For the OAWG and pulse compression applications that follow, we shape six lines of the signal comb used for DCS. The waveform monitor [red box in Fig.~\ref{fig2}(a-iii)] is used to measure the optical spectrum and waveform in real time. The signal comb exiting the chip is split by a fiber coupler; half of the light is sent for DCS phase measurement while the other half is sent for temporal/spectral analysis using a $30$~GHz PD and $50$~GHz sampling oscilloscope and optical spectrum analyzer (OSA). 
Since the global phase of the output is unobservable in our application, we can automatically assign one of the six shaper phases to zero. Moreover, with the freedom to choose the time origin, any linear contribution to the spectral phase (equivalent to a group delay) can be ignored as well, allowing us to set a second phase shifter to zero. In the context of the tests below, this flexibility allows us to apply zero drive to channels 3 and 6 without loss of functionality.


As a demonstration of the precise phase control achieved in our system, we generate a set of waveforms on a $3$~GHz frequency grid from the six EO comb lines. 
The ideal waveforms we target, without factoring in our measurement system response, are presented in Fig.~\ref{fig3}(a,b,c-i) to show the fine temporal features. 
The temporal profile of our initial phase state, with channels tuned to a $3$~GHz grid and phase shifters driven to induce $\sim\pi$ phase, is shown in Fig.~\ref{fig3}(a-ii) (black dashed line).
The corresponding measured phases for this untuned case (resonators aligned to $3$~GHz grid and four phase shifters driven with $\sim \pi$)are plotted in Fig.~\ref{fig3}(a-iii), which shows phase variation across each channel (excluding the two phase reference channels that are always set to zeros).   
The phase deviations could be caused by variations in tuning efficiency for each phase shifter, variations in optical path length for each channel caused by fabrication, or the input comb's nonuniform spectral phase profile.
The waveform shows a significant difference after the shaper is programmed to achieve a uniform spectral phase, as shown in Fig.~\ref{fig3}(a-ii) (blue line).
We also include the ideal waveform (Fig.~\ref{fig3}(a-i)) convolved with our measurement system impulse response in Fig.~\ref{fig3}(a-ii) (orange trace), demonstrating excellent agreement.  
Next, we target a phase vector implementing the temporal Talbot effect~\cite{huang2008spectral} where the phase modulation imposed on each line causes the temporal waveform to exhibit a multiplied repetition rate. We choose a phase vector [above Fig.~\ref{fig3}(b-i)] such that the repetition rate is doubled.
The various results for this phase state are shown in Fig. \ref{fig3}(b-i,ii,iii). Finally, we program another state, which we term the forked state, with waveform and the phase vector shown in and above Fig.~\ref{fig3}(c-i). As before, the various results are shown in Fig.~\ref{fig3}(c-i,ii,iii). We note that the Talbot and forked states are more sensitive to phase accuracy than the compressed pulse. For these waveforms, unlike the compressed state, the phases on each line need to be fine-tuned to obtain the waveform closest to the theoretical prediction. This could be caused by an imperfect reference comb phase measurement and results in a target signal comb phase that is slightly different than the ideal signal comb phase [see Fig.~\ref{fig3}(b,c-iii)]. For all states, we see excellent agreement with the target in both the temporal waveform and measured spectral phase.

\begin{figure*}[!t]%
\centering
\includegraphics[width=0.9\textwidth]{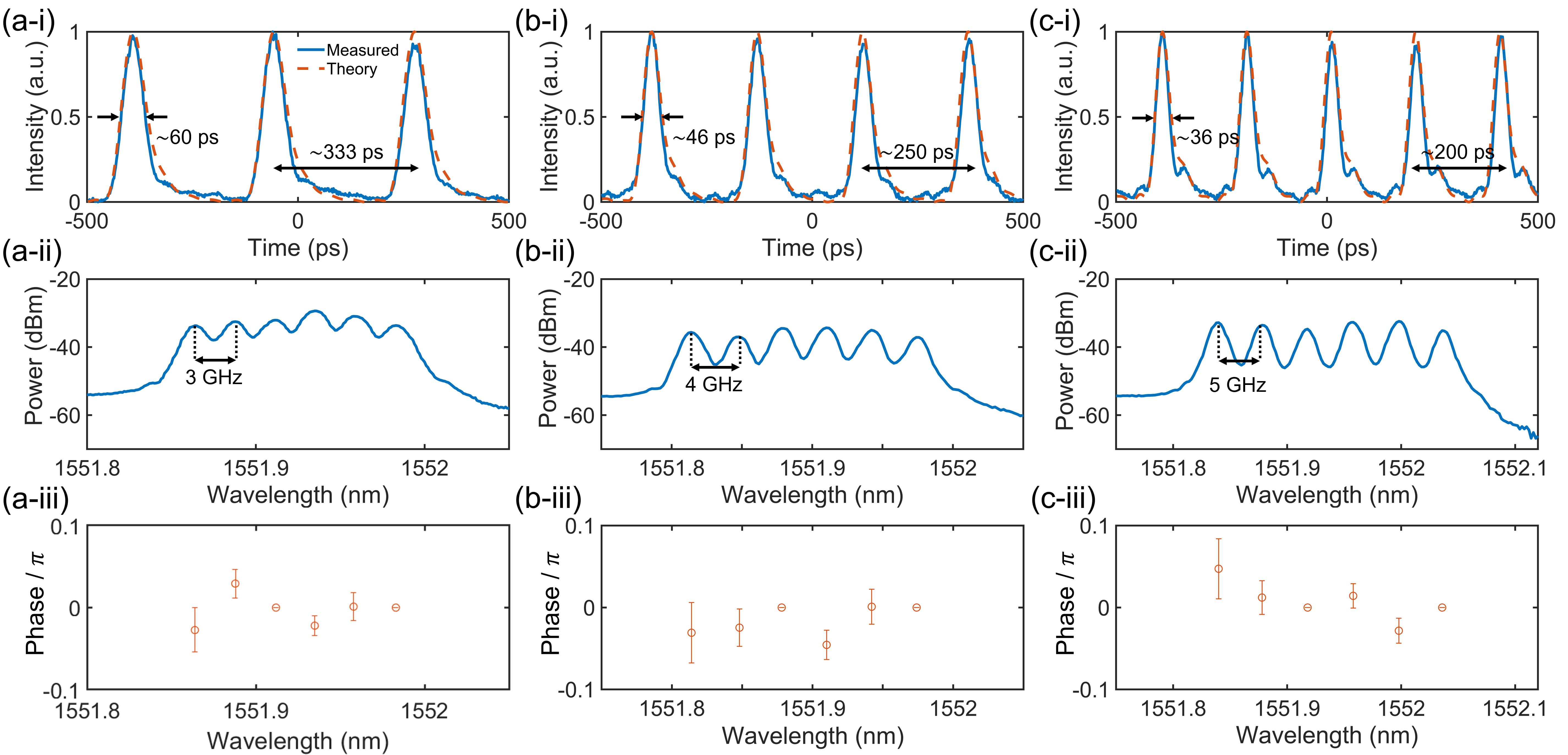}
\caption{\textbf{Pulse Compression} (a,b,c-i) Waveforms acquired from a $30$~GHz PD and $50$~GHz sampling oscilloscope (blue traces) compared with the simulated target waveform after convolution with our measurement system impulse response (red dashed traces), for $3$, $4$, and $5$~GHz channel spacing, respectively. (a,b,c-ii) OSA power spectral density of the EO comb lines coming out of the shaper after being programmed for compression on $3$, $4$, and $5$~GHz grids, respectively. The power measured here for each spectral line is taken into account in the theoretical trace in (a,b,c-i). (a,b,c-iii) The difference between measured and target phase vectors with error bars (standard deviation) from the measured phases, after programming for $3$, $4$, and $5$~GHz grids, respectively. 
}\label{fig4}
\end{figure*}

\noindent{\textbf{Pulse compression}}

We next show pulse compression for six-line combs with varying repetition rates of $f_{rep} = 3$, $4$, and $5$~GHz.
Once the routine converges, the measured output pulses in the time domain are shown for the $3$, $4$, and $5$~GHz repetition rates in Fig.~\ref{fig4}(a,b,c-i) (blue trace), respectively. The measured waveforms are plotted alongside the simulated ideal waveforms taking into account the response of our measurement system as well as the optical power of the comb lines in Fig.~\ref{fig4}(a,b,c-i) (orange trace). The measured power spectrum from the OSA (resolution $\sim 1.25$~GHz) of the output comb for each repetition rate is shown in Fig.~\ref{fig4}(a,b,c-ii). The apparent difference in maximum power between the lines is primarily due to the non-flat region of the EO comb lines sent into the chip.
Finally, the difference between the final measured spectral phases and the target phases are shown in Fig.~\ref{fig4}(a,b,c-iii).
We see good agreement between the measured and target temporal waveforms and spectral phases for each grid spacing, showcasing the tunability of our shaper.

\vspace{\baselineskip}
\noindent{\textbf{Discussion}}

\noindent Future shaper designs aim to reduce channel insertion loss (here $\sim 6$~dB) while maintaining similar or narrower channel linewidth (here $\sim 900$~MHz), and to increase the number of shaper channels. The addition of a substrate undercut, thermal trenching process~\cite{VanNiekerk2022,Rizzo2023}, the integration of phase change materials on SOI~\cite{Rios2022}, or using material platforms with intrinsic nonlinearities like LiNbO$_{3}$~\cite{Zhu:21} or AlGaAs~\cite{Chiles:19} could significantly reduce crosstalk effects and power consumption and simplify use of the shaper. 

Our measurement process can be enhanced on the MHS side by both using a lower repetition-rate MLFC or using multiple CW laser lines (a DCS approach). The first modification would result in a finer resolution measurement of the optical spectrum leading to more accurate channel alignment. The second would increase the measurable optical bandwidth allowing us to scale to larger channel numbers or spacings. Besides reducing the repetition rate of the MLFC, a spline or Lorentzian fit~\cite{Feng2012} could be applied to the MHS spectra to more accurately measure resonance frequencies.
The DCS measurement can easily be scaled to accommodate a larger number of shaper channels by increasing the number of components in the EO comb generators or adding components for nonlinear spectral broadening~\cite{Torres-Company2014}.
Fast PDs ($> 200$~GHz) on the SOI platform have been demonstrated~\cite{Lischke2021}, the integration of which could simplify the off-chip setup, while allowing for our tuning method to operate in a similar way. An FPGA, or monolithic electronic-photonic integration~\cite{Atabaki2018}, could replace oscilloscopes and embed electrical signal processing methods to reduce runtime. 
We note there are numerous other approaches for active control that could be useful for our system, for example involving dithering signals or photoconductive measurements~\cite{Padmaraju,Morichetti,Jayatilleka:19}.

A comparison between our SiP shaper and state-of-the-art alternatives is helpful: 
Commercial Fourier-transform pulse shapers today offer polarization independence and tunable channel bandwidths and amplitudes, yet are limited to $\sim 10$~GHz resolution and $\sim 5$~dB insertion loss (up to $2$~dB increase below $15$~GHz resolution)~\cite{coherentWaveshaper}. Bulk-optics Fourier shapers assembled in research laboratories have shown $< 400$~MHz resolution with an insertion loss of $\sim 15$~dB \cite{Willits:12,Lee2023}. On-chip Fourier-type shapers implementing AWGs have been shown in silica~\cite{fontaine200732}, InP~\cite{FONTAINE20113693,soares2011monolithic, metcalf2016integrated}, and Si$_3$N$_4$~\cite{Feng2017a} platforms with typical channel bandwidths and losses in the $10$s of GHz and $~3$-$10$~dB range, respectively. AWGs have also been used as de(multiplexers) in tandem with free-space imaging systems and LCoS modulators to realize shapers with a high resolution of $800$~MHz~\cite{rudnick2017sub}. Comparatively, non-Fourier shapers, based on Frequency Shifting Loops (FSL) and (or) nonlinear optical techniques, have been shown with very high resolution ($\sim 10$-$100$~MHz) and low losses \cite{DeChatellus2018,schnebelin2019programmable,redding2022high}. However, the demonstrations in~\cite{DeChatellus2018,schnebelin2019programmable} are limited to generating complex waveforms based on a CW seed laser and cannot shape a broadband optical field. Nonlinear methods based on Stimulated Brillouin Scattering (SBS)~\cite{redding2022high} are capable of operating on an arbitrary optical field, yet they rely on Brillouin signals which become prohibitively weak as the power of the input field decreases, as is the case for quantum light. In contrast, our shaper is linear and is equally effective for the quantum light~\cite{Nussbaum:22,lu2022high}. 
Furthermore, our SiP shaper has, to the best of our knowledge, the highest resolution (channel linewidth $~900$~MHz) of any on-chip Fourier-transform pulse shaper to date. Combining our chip with advanced edge coupler designs~\cite{mu2020edge} and photonic wirebonding techniques~\cite{blaicher2020hybrid}, and fiber-to-fiber losses of our SiP shaper could be comparable to the commercial offering. 

In conclusion, here we demonstrate a SiP spectral shaper with microresonator filter banks designed beyond the single-mode regime to achieve a high spectral resolution of $\sim 900$~MHz, a significant improvement over the bulk commercial counterpart with $\sim$~10~GHz resolution. In addition, we present a novel method for adaptively programming such shapers using common spectroscopy techniques and electrical signal processing methods. Our platform is well-suited for large-scale integration to produce the next generation of high resolution pulse shapers.


\vspace{\baselineskip}
\noindent\textbf{Methods}

\noindent \small \textbf{Multi-Heterodyne Spectroscopy.}
We spectrally filter a portion of a stabilized mode-locked frequency comb (Menlo Systems) with a $250$~MHz repetition rate using a $100$~GHz DWDM filter and send it through the bus waveguides in the shaper, as indicated by the brown arrows in Fig. \ref{fig2}(a-i). 
The output comb with amplitude-modified lines is then combined with a CW laser fixed to emit at a wavelength near the targeted channel frequencies and fed to a PD to produce heterodyne beat notes. Erbium-doped fiber amplifiers (EDFAs) are used to amplify the optical signal prior to photodetection, along with DWDM filters to suppress amplified spontaneous emission (ASE). We use two $50$~GHz PDs (u2tXPDV2020R, Coherent BPDV2120R) connected to a four-channel $30$~GHz oscilloscope ($80$~GSPS, Keysight DSA93004L) to detect these beat notes. An FFT is performed on the oscilloscope after a $2 \mu$s acquisition and transferred via USB to a PC for processing. A single data transfer takes on the order of $\sim 10s$ of ms. RF tones are processed and resonance frequencies are identified using peak finding methods. 


\noindent \small \textbf{Dual-Comb Spectroscopy.}
Signal and reference EO combs are generated by cascading a single intensity (JDS Uniphase, EOspace) and phase modulator (EOspace). Each comb generator is driven with an independent signal generator (Hittite HMC-T2100) with a repetition rate difference of $\Delta f_{rep} = 50$~MHz. The signal generators are synchronized to a common clock ($10$~MHz). The intensity modulators are driven at an RF power that can produce a square-like pulse shape (amplitude of $V_{\pi}/2$)~\cite{Wu2010}. Approximately 26 dBm of RF power is provided to the phase modulator for each comb, leading to broadband combs spreading $\sim$ 7 lines on each side of the pump (see Supplemental Information). RF amplifiers are used before the modulators to obtain the appropriate RF power levels. An RF phase shifter is included in each EO comb to align the phases between the intensity modulator and phase modulator. To ensure similar RF power levels for the 3, 4 \& 5 GHz EO combs, and to align the phases between intensity and phase modulators, the detailed EO comb configurations (the number of RF amplifiers and the position of phase shifter) are different for each repetition rate. This is not a limitation to the method, but a requirement based on the available RF equipment. 

The $2^{\rm nd}$ - $7^{\rm th}$ signal comb lines at the higher wavelength side of the pump are sent to the chip 
The signal comb going through the shaper and the MHS frequency comb going through a loopback are coupled out of the chip through the same waveguide but are spectrally separated with a DWDM filter. A portion of the output signal comb is then combined with the reference comb, detected on a 5~GHz balanced PD (Thorlabs BDX3BA), and is measured with a 2~GHz real-time oscilloscope (Rohde \& Schwarz RTO1024). EDFAs are used to amplify the optical signal prior to photodetection, along with DWDM filters to suppress ASE. The scope acquires a 10~$\mu$s signal, consisting of 500 interferogram periods (one interferogram period is $1/ \Delta f_{rep}$) between the two combs. The signal generator synchronization signal is used to trigger the oscilloscope to minimize the timing jitter between each measurement. FFT processing of the acquired signal provides the desired phase values. To achieve an accurate phase measurement, we process 16 acquisitions for each DCS measurement to obtain a mean and standard deviation of the extracted phases. A linear phase reference is necessary as acquisitions show slight timing jitter between measurements, affecting the mean and standard deviation of the phases. The error bars in the measured phase plots are the standard deviation, which is usually less than 0.1 rad and is not accepted unless $<$~0.15 rad. We note that the error bar of the 100 MHz-beat ($\sim$~0.1 rad, channel 1) is slightly higher than the other beats ($\sim$0.08 rad for 150 MHz-beat, channel 2, and $\sim$0.05 rad for the other two beats, channels 4 and 5) possibly due to the low-frequency amplifier noise around $<$~100 MHz (See Fig.~S6(c) in the Supplementary Information).


\noindent \small \textbf{Control Routine}
The flowchart of the routine is shown in Fig.~\ref{fig2}(b).
Resonators are tuned within tolerance when the error signal, the difference between measured resonance frequencies $f$ and target frequencies $F$, satisfies $ Err_f = \lvert F-f \rvert < 250 \times 10^6$~Hz. Phases are tuned within tolerance when the error signal, the difference between the signal comb phases $\phi_{sig}$ and target phases $\phi_{target}$, satisfies $ Err_{\phi} = \lvert \phi_{target}-\phi_{sig} \rvert < 0.1$~radians.
The DCS measured phase vector is $\phi_{DCS} = \phi_{sig} - \phi_{ref}$ (where $\phi_{sig}$ and $\phi_{ref}$ are absolute phase vectors of the signal and reference combs, respectively). $\phi_{sig}$ can be controlled arbitrarily when an accurate value of $\phi_{ref}$ is supplied (see Reference Comb Phase Measurement section in Methods for measurement details of $\phi_{ref}$).
As an example, for the pulse compression waveforms $\phi_{sig} = 0$, which means the DCS measured phase $\phi_{DCS} = -\phi_{ref}$. We approximate $\phi_{ref}$ with quadratic frequency dependence and measure the quadratic coefficient to be $\sim 0.009$~rad$/$GHz$^{2}$.
Microheaters in the resonators and phase shifters are driven with linear steps in power proportional to the respective error signals $Err_f$ and $Err_{\phi}$. The multichannel DC power supply (Nicslab XPOW-40AX-CCvCV-U-V12) updates drives in a serial fashion with a $\sim$~ms setting time. Practically, the control routine takes a couple hundred iterations or just a few minutes to converge to the desired state (see Supplementary Information for an example). However, in scenarios where phases require fine-tuning like that described for the states in Fig.~\ref{fig3}, the control sequence can be run once to converge near to the target state, and additional control sequences can be run with this state from the previous iteration as the starting point. In these cases, convergence to the next state can be reached in a few seconds.

\noindent \small \textbf{Reference Comb Phase Measurement.}
We program the shaper to pass only two channels. Two lines from the reference comb are sent through the chip, detected by a $30$~GHz BPD (Thorlabs BDX3BA), and measured with a $50$~GHz sampling oscilloscope (HP 54120). The sampling oscilloscope is triggered by a signal derived from the signal generator (using a $\div 3$ divider) driving the reference comb. The relative phase between the two comb lines is measured from the beat note as a temporal delay~\cite{jiang2006optical}. By tuning the pump laser wavelength in increments of the comb repetition rate, the relative phase difference between each adjacent pair of comb lines can be measured, and the spectral phase of the comb can be computed. A quadratic spectral phase provides a good fit to the measured values (see Supplementary Information). 

\vspace{\baselineskip}
\noindent \textbf{Data Availability}

\noindent Data are available from the authors upon reasonable request.

\renewcommand{\bibfont}{\small}
\setlength{\bibsep}{0.0pt}
\bibliography{ms}

\vspace{\baselineskip}
\noindent \textbf{Funding}

\noindent This work was funded under NSF grant 2034019-ECCS.

\vspace{\baselineskip}
\noindent \textbf{Acknowledgements}

\noindent The authors would like to thank Jason D. McKinney and Joseph M. Lukens for reviewing the manuscript, Vivek Wankhade for helpful discussions, and Matthew van Niekerk and Stefan Preble for providing wirebonding materials. 

\vspace{\baselineskip}
\noindent \textbf{Author Contributions}

\noindent L.M.C. designed the SiP chip and conceived the idea for the control method. L.M.C. and K.W. carried out the experiments and drafted the manuscript. L.M.C., K.W., K.V.M., and S.F. contributed to the experimental setup. N.B.L. and A.M.W. initialized the project. A.M.W. supervised the work. All authors discussed the results and contributed to the writing of the manuscript.

\noindent

\vspace{\baselineskip}
\noindent \textbf{Conflicts of Interest}

\noindent The authors declare no conflicts of interest.

\vspace{\baselineskip}
\noindent \textbf{Additional information}

\noindent More experimental details can be found in the Supplemental document. 

\end{document}


\title[Supplementary Information]{Silicon Photonic Microresonator-Based High-Resolution Line-by-Line Pulse Shaping: Supplementary Information}


\author*[1]{\fnm{Lucas} \sur{M. Cohen}} \email{\small cohen26@purdue.edu}

\author*[1]{\fnm{Kaiyi} \sur{Wu}} \email{\small wu1871@purdue.edu} 

\author[1]{\fnm{Karthik} \sur{V. Myilswamy}}

\author[1]{\fnm{Saleha} \sur{Fatema}}

\author[2]{\fnm{Navin} \sur{B. Lingaraju}}

\author[1]{\fnm{Andrew} \sur{M. Weiner}}

\affil[1]{\small\orgdiv{School of Electrical and Computer Engineering}, \orgname{Purdue University}, \orgaddress{\city{West Lafayette}, \postcode{47907}, \state{IN}, \country{USA}}}

\affil[2]{\small\orgdiv{The Johns Hopkins University Applied Physics Laboratory}, \orgaddress{\city{Laurel}, \postcode{20723}, \state{MD}, \country{USA}}}

\maketitle


\section{Transmission Spectra}

\begin{figure}[h]
    \centering
    \includegraphics[width=0.6\textwidth]{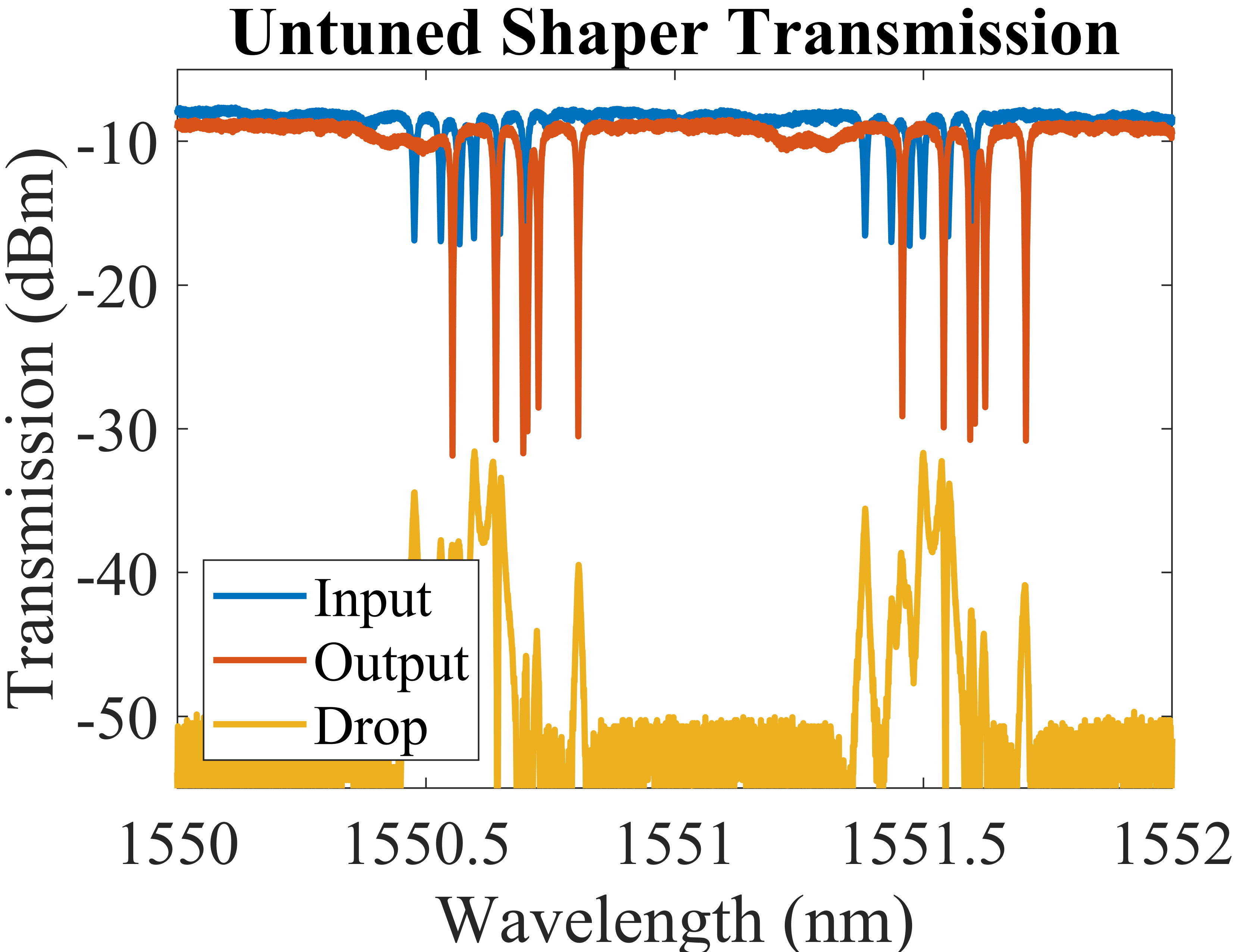}
    \caption{Transmission spectra of the as-fabricated spectral shaper. The blue and orange curves are the through port spectra for the input and output resonator banks, respectively. The yellow curve is the drop port response, where the light propagates through both resonators and phase shifter within each shaper channel.
    }
    \label{fig:untuned_shaper}
\end{figure}

A detailed description of the resonator design can be found in \cite{Cohen2022}. Compared to that work, this run results in a $\sim 2$~dB increase in insertion (drop port) loss. We attribute this to fabrication process differences between an active and passive MPW.
The shaper system is laid out with twelve identical resonators (two per channel). The resonance frequencies are initially scattered due to fabrication imperfections. 
Although already included in the main text, for completeness we show the as-fabricated transmission spectrum again here in Fig.~\ref{fig:untuned_shaper}. Here, the input optical power is $\sim 0$~dBm. A $\sim 3.5$~dB coupling loss per facet is measured. The transmission is taken via bus waveguides on both sides of the device. The as-fabricated shaper drop port response is measured when probing and monitoring the appropriate ports. Although designed to be identical, the resonators on the input and output bank exhibit different extinction ratios. 
This is probably due to different waveguide coupling strengths 
caused by fabrication differences. This phenomenon is also evident in the MHS spectra of Fig.~\ref{fig:MHS_5G}.

\newpage

\begin{figure}[t]
    \centering
    \includegraphics[width=\textwidth]{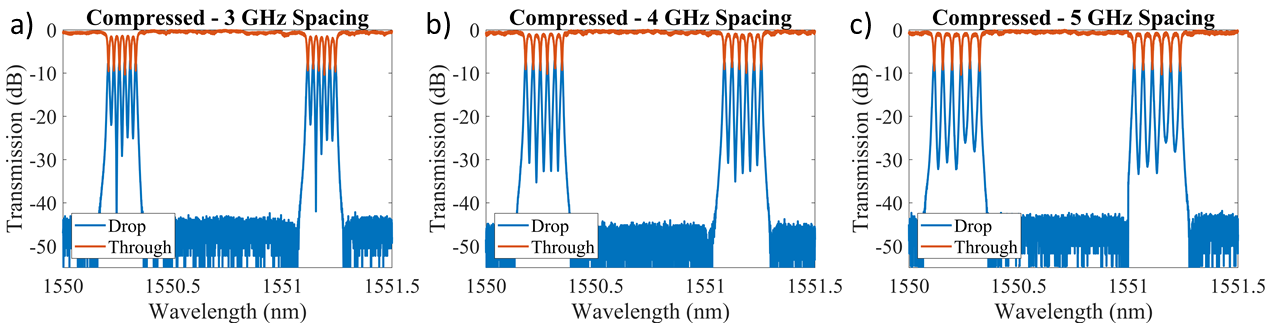}
    \caption{Transmission spectra to produce compressed optical pulses at (a) $3$, (b) $4$, and (c) $5$ GHz repetition rates. The lower-wavelength FSR is used for channel alignment with MHS and the higher-wavelength FSR is for DCS measurement and pulse compression.}
    \label{fig:tuned_various_grids}
\end{figure}

\begin{figure}[t]
    \centering
    \includegraphics[width=\textwidth]{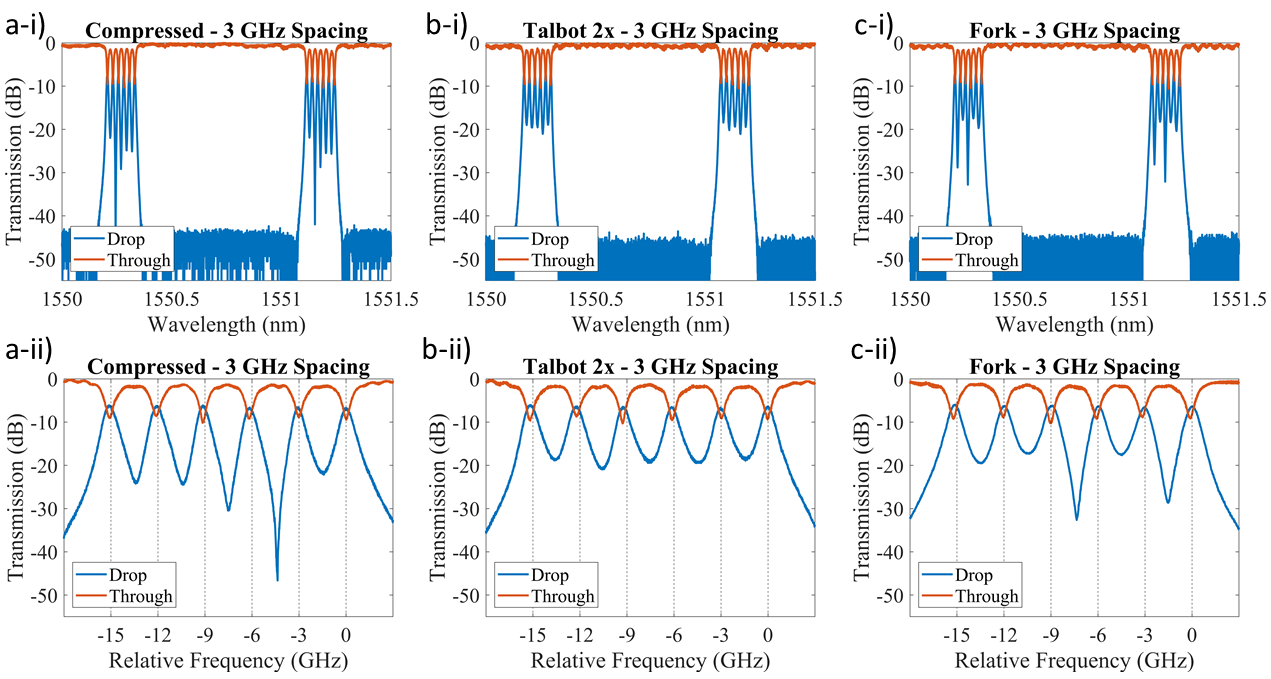}
    \caption{Transmission spectra and a zoom-in of the channel tuning FSR for (a-i,ii) the compressed state , (a-i,ii) the Talbot state, and (c-i,ii) the forked state at $3$~GHz repetition rate. In the bottom row, the vertical dashed black lines are on the target frequency positions at a $3$~GHz grid.}
    \label{fig:3G_states}
\end{figure}

\noindent The transmission spectra for the shaper tuned to phase-compensate EO comb lines of varying repetition rates ($3$, $4$, and $5$~GHz) are shown in Fig.~\ref{fig:tuned_various_grids} for each of the $3$, $4$, and $5$~GHz frequency grids. Next, all the spectra for the various $3$~GHz states are shown in Fig.~\ref{fig:3G_states}. The laser sweep spectra show good agreement with the target response. 

\pagebreak

\section{Multi-Heterodyne Spectroscopy Measurement}

\begin{figure}[b]
    \centering
    \includegraphics[width=\textwidth]{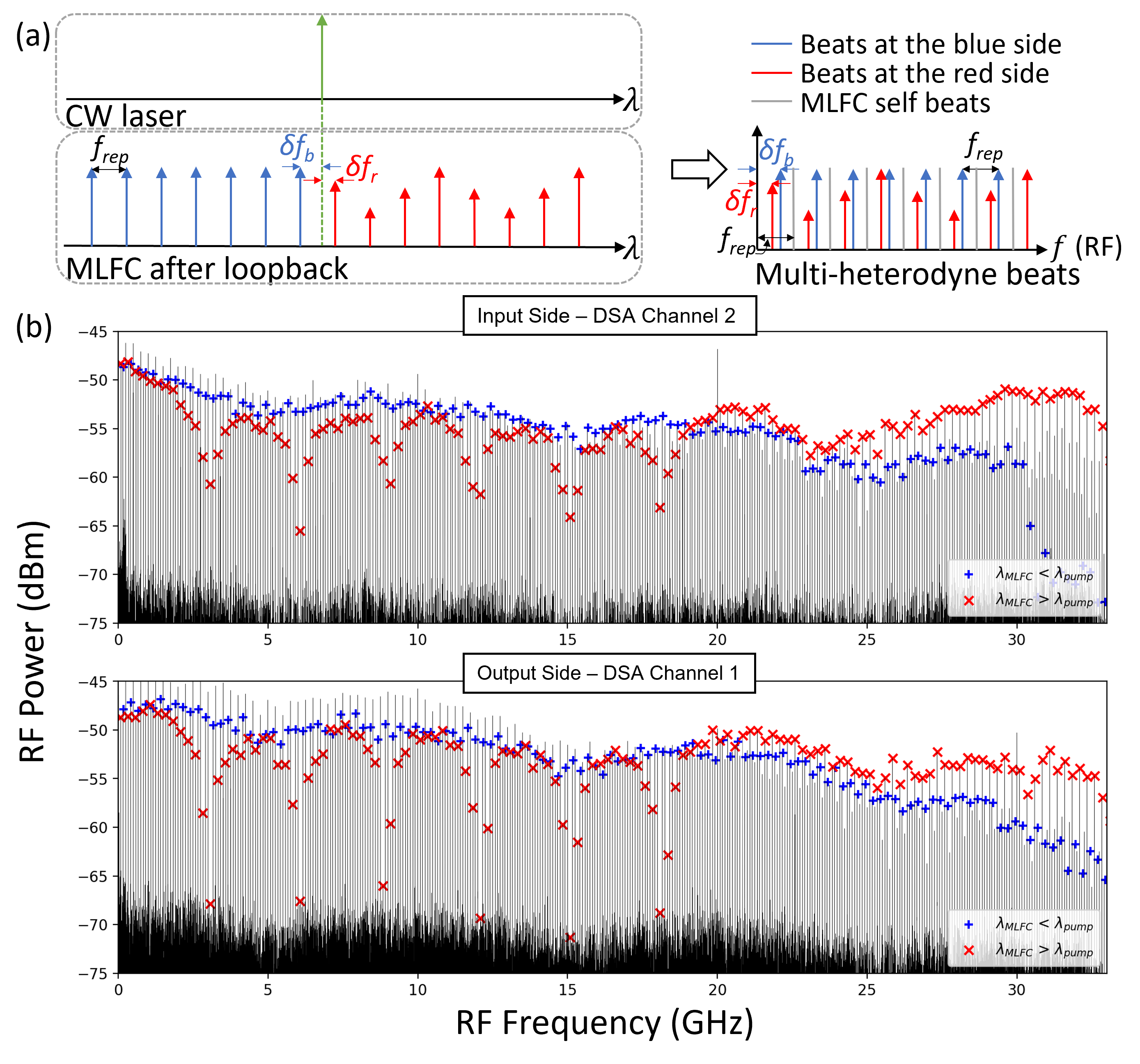}
    \caption{(a) Illustration of the MHS method. (b) RF spectra obtained from the Digital Signal Analyzer (DSA) showing RF tones modulated by the input and output resonator banks. The set of blue "$+$"s and red "$\times$"s correspond to beat notes between the fixed wavelength laser and the MLFC lines at lower and higher wavelengths than the laser, respectively. The tones without markers correspond to self-beats from the MLFC.}
    \label{fig:MHS_5G}
\end{figure}

Figure~\ref{fig:MHS_5G}(a) illustrates the MHS method. A CW laser at a fixed wavelength is combined with the spectrally modified (sent through bus waveguides on the chip) mode-locked frequency comb (MLFC) to produce RF tones spaced at the comb's repetition rate $f_{\rm rep}$ (as described in the text). If the laser wavelength is slightly detuned from the center of two MLFC lines, two sets of RF beat notes will appear at a different starting frequency ($\delta f_{\rm b}$ and $\delta f_{\rm r}$ as illustrated) and will repeat with the repetition rate of the MLFC \cite{urabe2013multiheterodyne}. These beat pairs are from contributions from MLFC lines at the red (start at $\delta f_{\rm r}$ and correspond to higher wavelength) and blue (start at $\delta f_{\rm b}$ and correspond to lower wavelength) sides of the laser. Fig. {\ref{fig:MHS_5G}}(b) shows two RF spectra obtained from the MHS method. The upper (lower) spectrum is measured from the MLFC which is spectrally modified by the input (output) resonator bank. The two spectra are obtained simultaneously by acquiring the MHS signals on two different DSA channels (DSA channels 1 and 2, as denoted in the Figure titles) and performing an FFT on each channel. There are many RF tones shown; the peak RF power of the red and blue tones, beats produced from the laser and MLFC lines of higher and lower wavelength, are indicated by the "$\times$" and "$+$" markers, respectively, with corresponding colors. The peak RF powers of these tones are used to track the resonance frequencies. The transparent black colored lines show the measured RF spectrum, including the self-beats between lines of the MLFC. 
In Fig.~{\ref{fig:MHS_5G}}(b) all resonators in each bank are tuned to be resonant at a greater wavelength than the laser, leading to spectral features appearing only at red tones, and aligned via our control scheme at $3, 6, 9, 12, 15$, and $18$~GHz.



\pagebreak

\section{Dual-Comb Spectroscopy Measurement}

\begin{figure}[h]
    \centering
    \includegraphics[width=\textwidth]{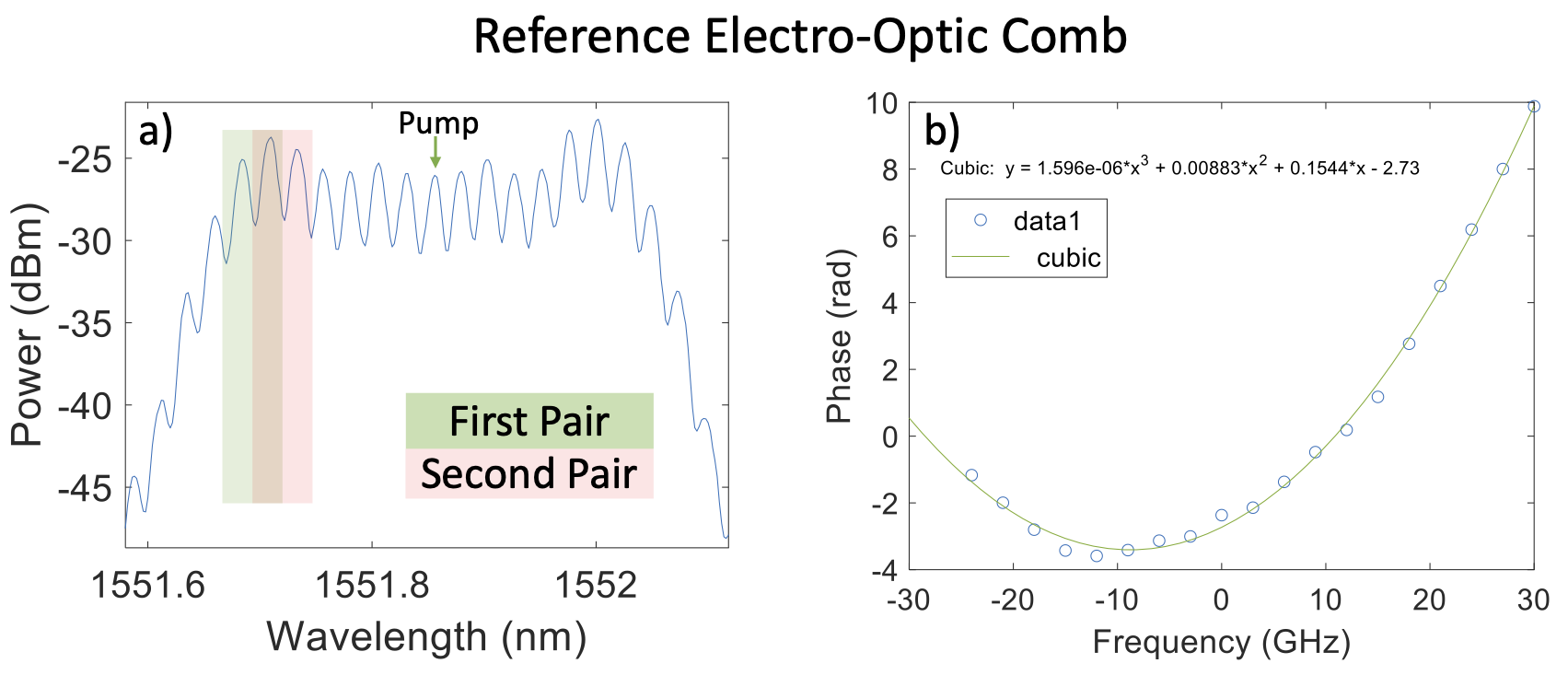}
    \caption{(a) Power spectral density for the reference EO comb. Green and red highlighted regions indicate comb line pairs sent into the shaper for time-resolved spectral phase measurement. (b) Unwrapped reference EO comb spectral phase with cubic polynomial fit. In the fitted formula, x is the x-axis (frequency in GHz unit) and y is the y-axis (unwrapped phase in rad unit)}
    \label{fig:Reference_Comb}
\end{figure}

The reference EO comb spectrum, measured from a $1.25$~GHz resolution OSA, is shown in Fig.~\ref{fig:Reference_Comb}(a). We first measure the phase of the reference comb; 
pairs of adjacent comb lines are sent into the SiP spectral shaper programmed to transmit two channels at the comb spacing. The light exiting the shaper then beats on a PD to produce beat notes with a delay indicating the relative phase between the comb lines \cite{jiang2006optical}. In this way, the reference comb phase can be measured as shown in Fig.~\ref{fig:Reference_Comb}(b). A cubic fit is applied to the measured data which indicates the dominant spectral phase contribution is quadratic (linear phase corresponds to a time delay and is ignored). We use this quadratic coefficient to represent the reference comb phase.

\begin{figure}[h]
    \centering
    \includegraphics[width=\textwidth]{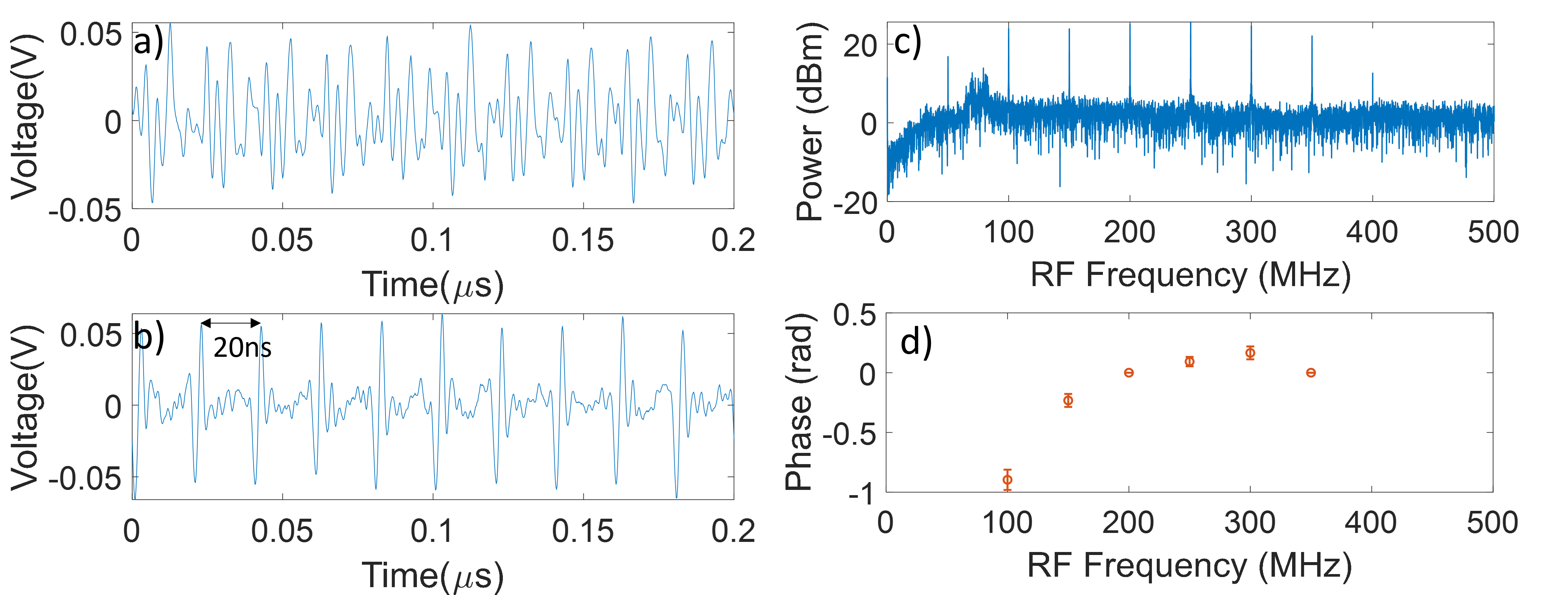}
        \caption{Time traces containing nine interferogram periods for the (a) untuned and (b) tuned $3$~GHz compressed state. The repetition rate difference of $\delta f = 50$~MHz sets the interferogram repetition rate. (c) Fast Fourier Transform (FFT) spectrum from the time trace in (b). Beat notes from $100$ to $350$~MHz are the primary six lines. The tones at $50$ and $400$~MHz are from crosstalk from the first and sixth shaper channel, respectively, and are attenuated $> 10$~dB with respect to the primary tones. (d) Retrieved shaper phases from the FFT trace in (b) with error bars accumulated from the acquisition of 16 time traces on the oscilloscope when the shaper is tuned to compress the $3$~GHz comb lines.}
    \label{fig:DCS_tuned}
\end{figure}

\noindent The reference comb spectral phase is used to target a measured phase vector via DCS. A $200$~ns beat trace (nine interferograms) between the signal comb exiting the chip and the reference comb with the phase state untuned and tuned to pulse state are shown in Figs.~\ref{fig:DCS_tuned}(a) and (b), respectively. The signal and reference combs have repetition rates of $3$ and $3.05$~GHz, respectively. After an FFT, the beat notes are apparent (Fig.~\ref{fig:DCS_tuned}(c)) and the beat note phase can be used to measure the applied shaper phase. Figure~\ref{fig:DCS_tuned}(b) shows that the interferograms become more pulse-like when the signal comb is compressed.  

\pagebreak

\pagebreak

\section{Control Scheme}

\begin{figure}[h]
    \centering
    \includegraphics[width=\textwidth]{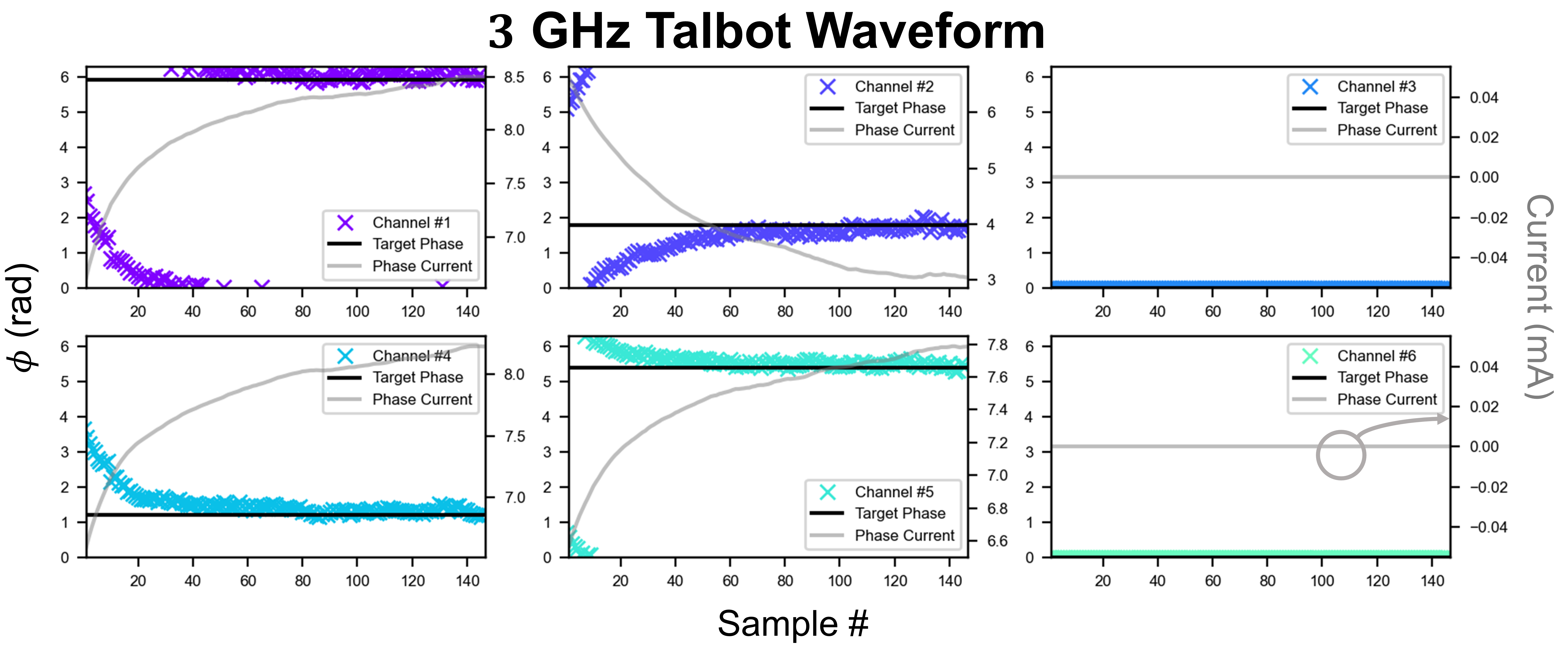}
    \caption{Measured phase (left y-axis, colored $\times$s) and applied phase shifter current (gray, right y-axis) versus sample number for all modes of the six-channel shaper for  $2 \times$~Talbot state on a $3$~GHz grid. Channel $3$ and $6$ are used as a linear and constant phase reference, respectively.}
    \label{fig:talbot_phase_evolution}
\end{figure}

\begin{figure}[h]
    \centering
    \includegraphics[width=\textwidth]{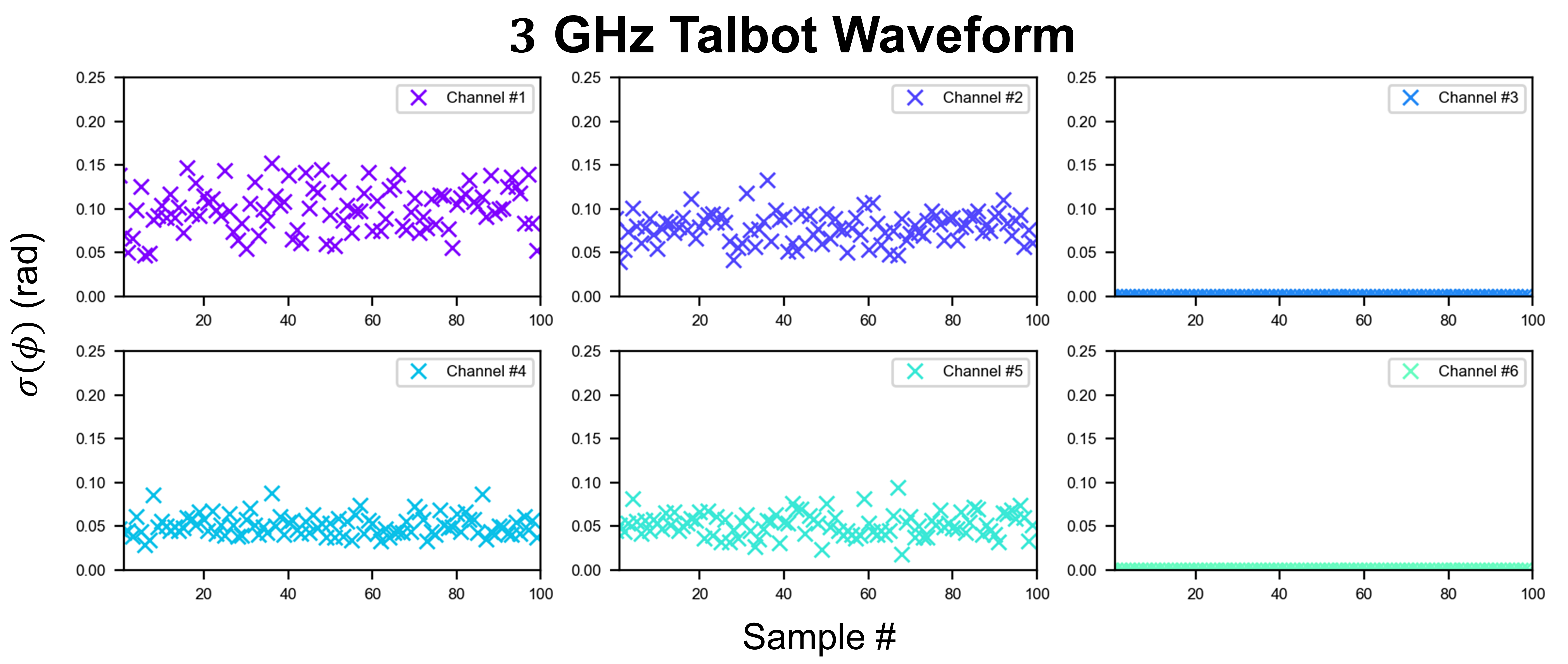}
    \caption{Standard deviation of measured phase versus sample number for all modes of the six-channel shaper. Channel $3$ and $6$ are used as a linear and constant phase reference, respectively. $16$ phase measurements are made via DCS to accumulate these statistics.}
    \label{fig:talbot_phase_std}
\end{figure}

We present the evolution of the phase on each line when running the control algorithm in Fig.~\ref{fig:talbot_phase_evolution} using the $2 \times$~Talbot state on a $3$~GHz grid as an example. The standard deviations are shown in Fig.~\ref{fig:talbot_phase_std}. The phase measurement data here are retrieved only at the iterations of the control routine when resonators are properly aligned (within tolerance). In this example, the total number of iterations is $\sim 200$. Each resonance frequency tuning iteration takes $\sim 100$~ms -- primarily limited by the processing speed, data transfer time, and the driving electronics update rate. Each phase measurement/tuning iteration takes an additional couple of seconds. This is mainly due to acquiring 16 time-domain waveforms and performing an FFT and additional data processing. On average, the entire control process takes a few minutes, depending on the initial state. However, in principle, both DCS and MHS data can be acquired at MHz rates. No effort was made to further reduce the control runtime. 
In scenarios where the desired shaper state is similar to the currently programmed state, like that described for the fine-tuning of states in Fig. 3 of the main text, the control runtime is drastically reduced. In the cases of the main text, convergence to the next state was reached in seconds.



Phase measurements are only accepted when the standard deviation is $< 0.15$ radians. The RF amplifier used to boost the electronic signal has a strong low-frequency noise causing Mode $1$ and $2$ to have slightly higher standard deviation than the others. Acquiring a longer time trace via DCS (and therefore averaging more interferograms) would reduce the standard deviation \cite{Duran2015}. The applied power to each phase shifter is incremented proportionally to the difference between the measured and target phase. This leads to the asymptotic convergence seen in the figure.

\renewcommand{\bibfont}{\small}
\setlength{\bibsep}{0.0pt}
